# Confinement Effect Emission from Infiltrated ZnO in PS-b-PMMA Nanostructures

Paris L. Blaisdell-Pijuan[1,2], David J. Gosztola[3], Angel Yanguas-Gil[3], Jiaxing Ren[2], Paul Nealey[2], Xuedan Ma[3], Stephen Gray[3], and Leonidas E. Ocola[3]

[1]California State University, Fullerton, 800 North State College Boulevard, Fullerton, California 92831, United States
[2]University of Chicago, 5801 South Ellis Avenue, Chicago, Illinois 60637, United States
[3]Argonne National Laboratory, 9700 South Cass Avenue, Argonne, Illinois 60439, United States

**ABSTRACT**

We have characterized the growth of ZnO using sequential infiltration synthesis (SiS) on PS-b-PMMA block copolymers (BCP) of spherical and cylindrical sub-20nm morphologies and studied how the photoluminescence of these nanostructures varies per its seed layer. Investigation of these structures was done using atomic force microscopy (AFM), spectrofluorometry, Raman spectroscopy, and scanning electron microscopy (SEM). We report blue-shifted photoemission at 335 nm (3.70 eV), suggesting quantum confinement effects. This UV-photoluminescence can be translated into emitter sizes of roughly 1.5nm in radius. Furthermore, samples of ZnO prepared with an alumina seed layer showed additional defect state photoemission at 470 nm and 520 nm for spherical and cylindrical BCP morphologies, respectively. Defect photoemission was not observed in samples prepared without a seed layer. Raman and EXAFS data suggest lack of long range order between the ZnO nanostructures during early stages of infiltrated ZnO growth and therefore supports the blue shift emission is due to confinement. Our work demonstrates that ZnO nanostructures grown on PS-b-PMMA via infiltration are advantageous in uniformity and size, and exhibit unique fluorescence properties. These observations suggest that infiltrated ZnO in PS-b-PMMA nanostructures lends itself to a new regime of applications in photonics and quantum materials.

## 1. Introduction

Zinc Oxide (ZnO) remains one of the most extensively studied metal oxides for its applications in photonics, solid state devices, gas sensors, biosensors and more (**Yohn-Zhe 2009; Fang 2011; Zhu 2017;**). Incorporation of ZnO in these devices is typically done in the form of thin films, quantum dots, and nanostructures, via deposition and chemical synthesis. While ZnO grown in this manner has been thoroughly characterized, ZnO developed in composites remains understudied. Recent efforts incorporating ZnO into polymethyl-methacrylate (PMMA) have shown an increase in photoemission suggesting enhancement of optical properties via the ZnO-polymer bond **(Ocola 2016, Ocola 2017)**. This makes ZnO-PMMA nanostructures an attractive target for photonic, and sensor applications.

Infiltration synthesis has gained attention for its applications to imaging, lithography, and thin films **(Barry 2017, Elam 2015, Nam, Peng Peng)**. It is a variant of atomic layer deposition (ALD) that operates at lower temperatures and for longer time periods, which allows for infiltration in polymers without any deformation of the polymer. Typically, a seed layer is used as a precursor to the target material. However, the effects of a seed layer for materials synthesized with infiltration lack detailed studies. To this end, we investigate the effects of using a seed layer in the growth of nanoscale ZnO structures. Due to the bonds between two different semiconductor metal oxides with different band gaps we expect the optical properties to vary between ZnO grown on different seed layers.

Recently, infiltration has shown promise towards developing a ZnO/PMMA composite (**Ocola 2016, Lorenzoni 2017, Ocola 2017**). Furthermore, PMMA is a known component of block copolymers (BCP), and has been incorporated with polystyrene (PS) to form PS-b-PMMA. This BCP separates the polystyrene and PMMA into distinct domains that create isolated PMMA structures. These structures come in multiple morphologies some of which are spheres or cylinders (**Park 2007**). This allows us to use infiltration on the BCP so that ZnO nanostructures will form in the PMMA domains creating uniform ZnO nanostructures. In the current work, we will use both cylinder and spherical morphologies and study the effects of sub-nm confinement to the properties of infiltrated ZnO as a function of growth.

The infiltration process limits the deposition of molecules on a surface to one atom at a time. We use water ($H_2O$) and diethylzinc (DEZ) as precursors for ZnO and $H_2O$ and trimethylaluminum (TMA) as precursors for the aluminum oxide ($Al_2O_3$) for the seed. A detailed description of the growth of ZnO in PMMA thin films can be found in Ref (**Ocola 2017**). The reaction for infiltration on ZnO occurs as follows:

1. First pulse with water: surface + $H_2O$ → surface–OH
2. React with DEZ: surface–OH + $C_2H_5$–Zn–$C_2H_5$ → surface–O–Zn–$C_2H_5$ + $C_2H_6$
3. Cleave with water: surface–O–Zn–$C_2H_5$ + $H_2O$ → surface–O–Zn–OH + $C_2H_6$

where "surface" means a seeded or unseeded inner wall of the PMMA free volume.

In this paper, we report on the characterization of the $Al_2O_3$ seed layer dependence for infiltration grown ZnO on PS-b-PMMA using atomic force microscopy (AFM), Raman scattering, scanning electron microscopy (SEM), and by measuring fluorescence.

We find that polystyrene phase of the BCP has a clear photoemission spectrum that interferes with ZnO fluorescence, and demonstrate the use of zero bias oxygen-plasma reactive ion etching to eliminate the interference of polystyrene emission without damaging the infiltrated PMMA phase. In addition, we investigated ways to reduce the amount of defect state emitters via annealing. Reports in the literature indicate that annealing ZnO thin films and ZnO nanostructures is directly correlated to the number of defect states present in the sample (**Jang 2010, Lu 2014**). These defect states have characteristic photoemission that has been used to produce single photons (**Morfa 2012, Choi 2015, Jungwirth 2015, Nietzke 2015**). Specifically, annealing at 500 °C has been used to observe single photon emission around 640nm (**Choi 2014**). Finally, we explore other means to create isolated nanoemitters by means of electron beam lithography and reactive ion etching, with high-level spatial positioning precision.

## 2. Results and Discussion

Growth of ZnO on PS-b-PMMA BCP was done using an atomic layer deposition tool configured for infiltration while using tetra methyl ammonium (TMA and diethyl zinc (DEZ) for aluminum and zinc deposition, respectively. Details can be found in the Methods section. Samples were prepared with an $Al_2O_3$ seed layer and without a seed layer. **Table 1** details all samples prepared and tested. We also tested samples with two cycles of $Al_2O_3$ as a seed layer. AFM, Photoluminescence (PL), and Raman spectroscopy measurements did not vary significantly between samples prepared with one or two cycles of $Al_2O_3$ seed layer.

| Spherical Morphology | | Cylindrical Morphology | |
|---|---|---|---|
| No Seed Layer | 1x [H$_2$O:TMA] Seed Layer | No Seed Layer | 1x [H$_2$O:TMA] Seed Layer |
| 1x [H$_2$O:DEZ] | 1x [H$_2$O:DEZ] | 1x [H$_2$O:DEZ] | 1x [H$_2$O:DEZ] |
| 2x [H$_2$O:DEZ] | 2x [H$_2$O:DEZ] | 2x [H$_2$O:DEZ] | 2x [H$_2$O:DEZ] |
| 3x [H$_2$O:DEZ] | 3x [H$_2$O:DEZ] | 3x [H$_2$O:DEZ] | 3x [H$_2$O:DEZ] |
| 4x [H$_2$O:DEZ] | 4x [H$_2$O:DEZ] | 4x [H$_2$O:DEZ] | 4x [H$_2$O:DEZ] |
| 5x [H$_2$O:DEZ] | 5x [H$_2$O:DEZ] | 5x [H$_2$O:DEZ] | 5x [H$_2$O:DEZ] |

**Table 1:** Table representation of the 20 samples prepared. For each of the two BCP morphologies two varying seed layers were prepared. For each of BCP morphology and seed layer, five samples with increasing cycles of ZnO were studied.

AFM of infiltrated BCP spheres tracked growth for successive cycles of ZnO. The topography of ZnO grown on the BCP with an alumina seed layer varied greatly from ZnO grown on the BCP without a seed layer. ZnO grown with a seed layer formed dimples in the centers of the sphere and began to grow up the edges while samples without seed layers formed in concentric spheres. **Figure 1a** depicts a film with PMMA BCP spheres infiltrated with one cycle of water and TMA, 1x [H$_2$O:TMA], and five cycles of water and diethylzinc, 5x [H$_2$O:DEZ], while **Figure 1b** depicts PMMA BCP spheres with 5x [H$_2$O:DEZ]. The presence of the aluminum oxide creates dimples in the nanostructures. There is no bridging visible in any samples. The highest structures occurred with three cycles of ZnO. We postulate that this is correlated with peak ZnO photoemission, as discussed later.

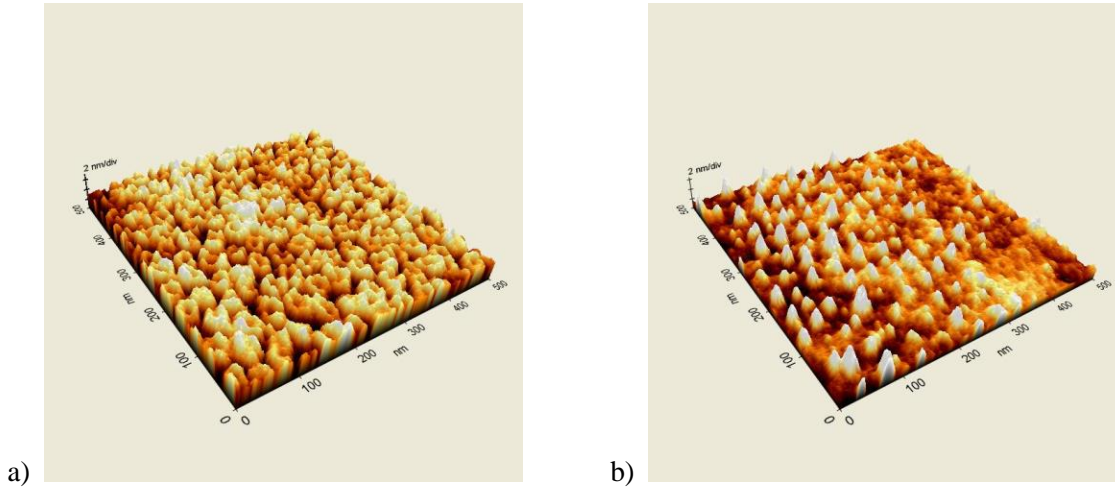

a)                  b)

**Figure 1**: Atomic force microscopy (AFM) image of spherical BCP with (a) 1x [H2O:TMA] and 5x [H2O:DEZ], and (b) 5x [H2O:DEZ]. Samples grown with an alumina seed layer exhibited dimples and grew higher, quicker. Samples grown without a seed layer formed concentric circles and remained relatively uniform.

Removal of the PS phase background was tested using two different methods. Two samples of cylindrical BCP were prepared with one Al$_2$O$_3$ seed layer and three cycles of ZnO. One sample was etched with UV-ozone for 10 min, while the other sample underwent 15s of reactive ion etching with zero biased oxygen-plasma. SEM of both samples was taken before and after the etching process, as shown in **Figure 2**. The sample etched with oxygen-plasma had significantly more PS removed without damage to the nanostructure PMMA/ZnO domains. To test the upper limit of the oxygen-plasma etch we administered another 25s of non-biased oxygen-plasma. SEM inspection after the additional etch did not reveal any observable damage to the infiltrated sphere or cylinders. From this test all samples were administered 40s of the non-biased oxygen-plasma etching process to remove the PS phase after infiltration. When running this same process on infiltrated BCP samples without the $Al_2O_3$ seed layer we find that the ZnO agglomerated in clusters, **Figure 3.** This implies that the infiltrated $Al_2O_3$ seed layer provides a stable and uniform scaffold (**Segal 2015**) for the synthesis of infiltrated ZnO and that the ZnO indeed is not a connected network, but simply isolated nucleated molecules at the initial stages of growth.

Raman spectroscopy measurements with a 325 nm excitation source were taken of all samples to determine the presence of any longitudinal optical (LO) phonons as a function of growth. A Raman peak was observed at 590 $cm^{-1}$ for the BCP with cylindrical morphology with 1x [$H_2O$:TMA] and 5x [$H_2O$:DEZ], indicating the presence of LO phonons. For all other samples tested no Raman peak was observed. These measurements indicate the absence of long range order between ZnO nanostructures. This suggests that five layers of ZnO may be an upper threshold for isolation of these spheres. Our data indicates that infiltration grown ZnO on BCP takes longer to form thin films than on PMMA alone (**Ocola 2017**). This is consistent with a recent AFM study done on the PS-b-PMMA for growth of $Al_2O_3$, which reported bridging of structures after five layers of $Al_2O_3$ (**Lorenzoni 2017**). **Figure 4** details the Raman excitation of infiltration grown ZnO on cylindrical PS-b-PMMA with one TMA seed layer as a function of growth. The lack of Raman peaks at 590 $cm^{-1}$ strongly suggest that the formed ZnO are a collection of isolated molecular emitters of the form: $Al_1O_{m+2}Zn_m$, where *m (=1 to 4)* is indicative of how many cycles were used. Recent EXAFS data on PMMA thin films corroborate this suggestion with conclusive evidence that at 2 infiltration ZnO cycles the Zn atoms only have one Zn atom nearest neighbor, **Figure 5**.

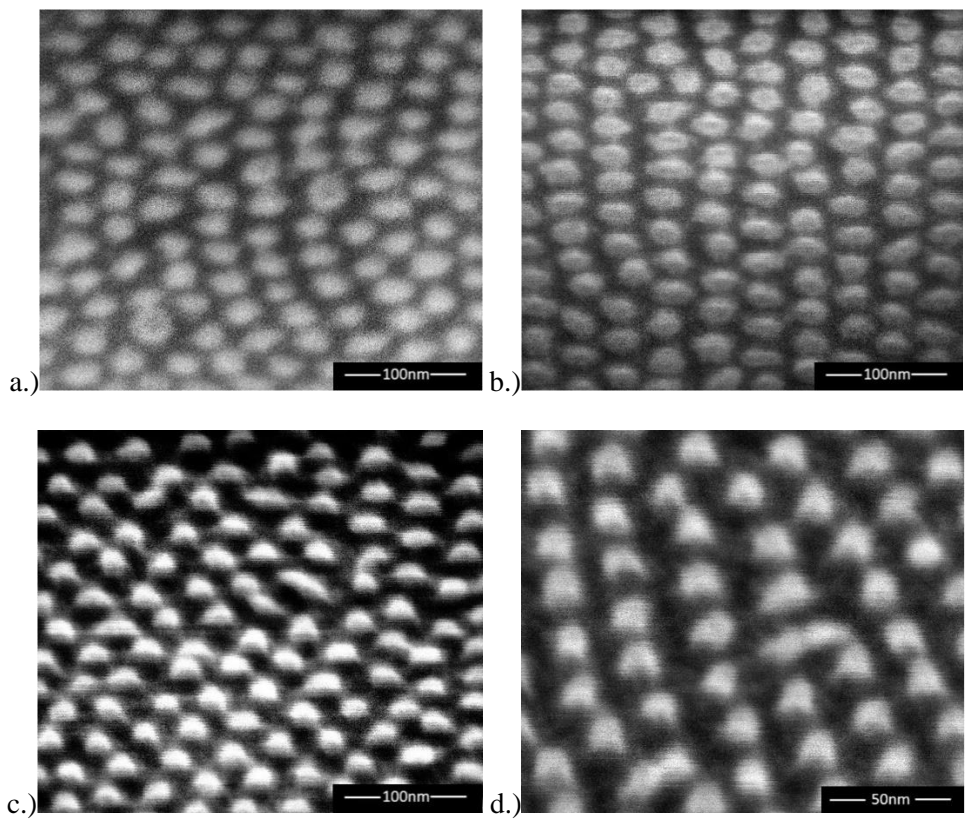

**Figure 2:** SEM images of cylindrical PS-b-PMMA infiltrated with 1x [$H_2O$:TMA] and 3x [$H_2O$:DEZ] (a) before any etching, (b) after etching with UV-ozone for 10 min., (c) after etching with zero-biased oxygen-plasma for 15s, and (d) after etching with zero-biased oxygen-plasma for 40s.

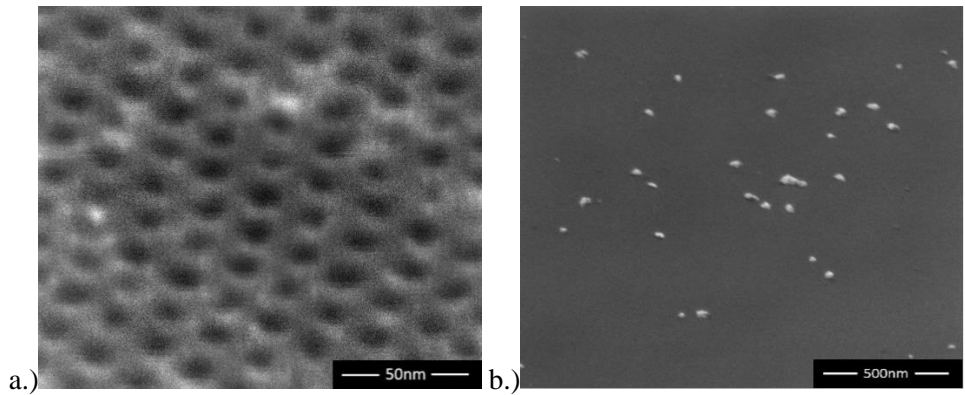

**Figure 3:** SEM images of cylindrical PS-b-PMMA infiltrated with 3x [$H_2O$:DEZ] (a) before any etching, and (b) after etching with non-biased oxygen-plasma for 40s

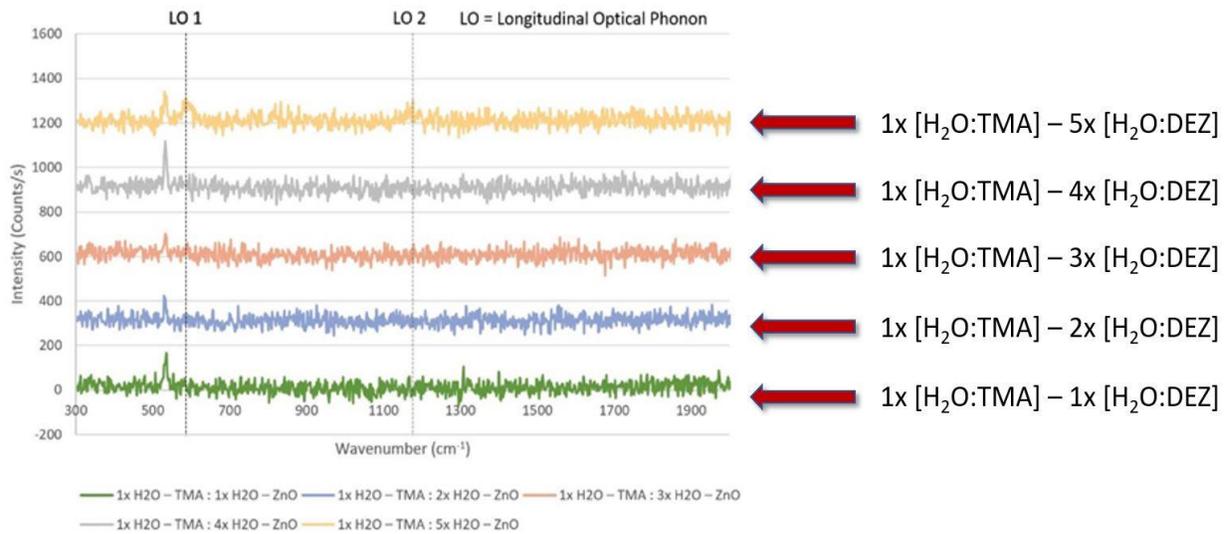

**Figure 4:** Raman Scattering of ZnO infiltrated PS-b-PMMA Cylinders with 1x TMA seed layer. Peaks at 530 cm$^{-1}$ indicate Raman peaks due to the SiO2 background. The first and second longitudinal optical phonons are located at 590 cm$^{-1}$ and 1185 cm$^{-1}$, respectively. The absence of these Raman peaks in samples with less than five cycles of ZnO indicate isolated nanostructures, and the absence of thin film. This means we can implement up to four cycles of ZnO and still have isolated nanostructures without bridging.

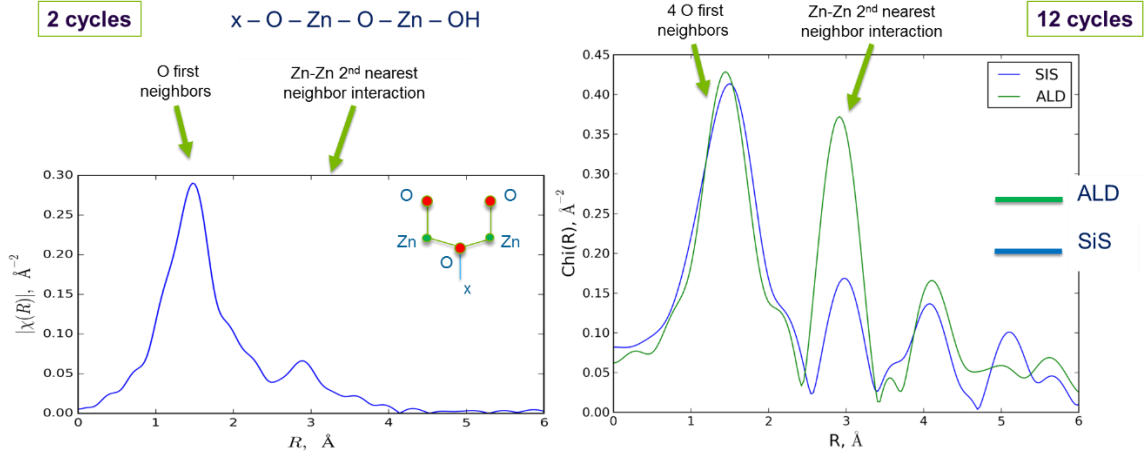

**Figure 5**. EXAFS data on ZnO grown in thin PMMA films comparing two cycles of infiltration ZnO with "bulk like" infiltration ZnO at 12 cycles and also a thin film of ZnO grown by regular ALD. Each peak indicates coordination occupancy by neighboring atoms. At two cycles only three oxygen first nearest neighbors and one zinc atom second nearest neighbors are found. At twelve cycles full coordination of four oxygen nearest neighbors match ALD ZnO and Zn second neighbors are less in infiltration ZnO than ALD ZnO indicating a diffuse or distorted ZnO structure.

Photoluminescence excitation mappings were taken to locate the peak excitation wavelength corresponding to peak emission. Excitation ranged from 250nm – 280nm in 3nm steps, using a 3nm slit width, and a 0.3s integration time. Emission spectra ranged between 320nm – 600nm with peak excitation and emission at 260nm and 335nm, respectively. A 305nm filter was placed in front of the detector to screen out any artifacts from reflected light. Samples without a seed layer showed no change in emission as a function of growth while samples with a seed layer showed additional emission at 470nm and 520nm for spherical and cylindrical BCP morphology, respectively. This late-growth photoemission at 470nm and 520nm is characteristic of ZnO defect state emission. Surprisingly, there are different defect states based on the morphology of the BCP. The photoluminescence of spherical and cylindrical BCP samples is given as a function of ZnO growth in **Figure 6** and **Figure 7**, respectively. In addition, the 335nm (3.7eV) photoluminescence present in all samples is blue-shifted from the 370nm (3.35 eV) emission of bulk ZnO, suggesting quantum confinement effects. For most growth variations, this emission is strongest after three cycles of ZnO. Photoemission energy has been shown to be directly related to emitter size **(Brus 1983, Kim 2002, Lin 2005, Dijken 2000)**. Brus has developed an equation that shows how the photoemission energy ($E^*$) is related to the radius (R) of the nanoparticle **(Brus 1984, Brus 1986)**. Employing the appropriate energy gap and effective masses for ZnO **(Brus 1984)**, this equation becomes

$$E^* = 3.44 + \frac{2.40241}{R^2} - \frac{0.70052}{R} \qquad , \qquad (1)$$

where E* is in eV and R is in nm. Equation (1) implies that a 3.7 eV emission corresponds to an emitter radius of roughly 1.5 nm – 2 nm. This result clearly shows that by controlling the growth conditions with infiltration in BCP nanostructures we can observe confinement effects in infiltrated ZnO. This is the first time this effect has been reported.

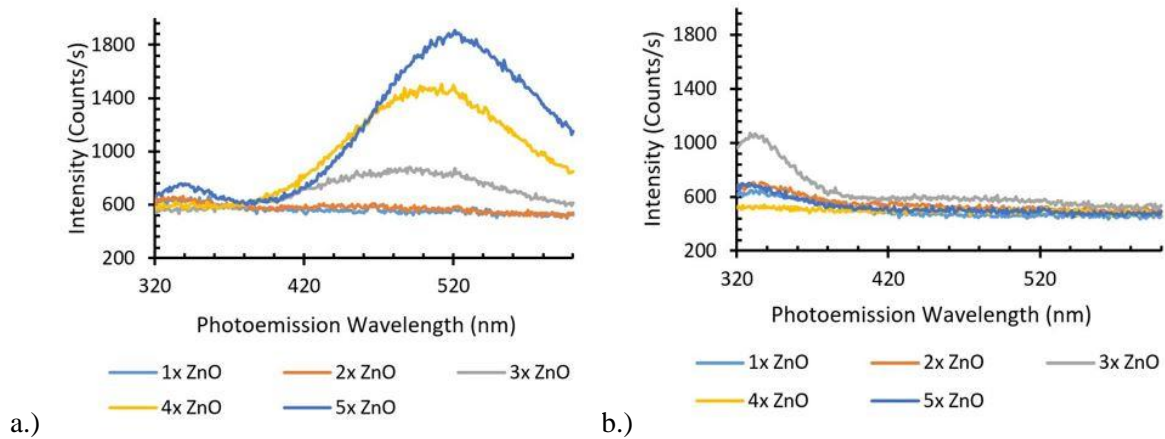

**Figure 6:** Photoluminescence of ZnO infiltrated PS-b-PMMA of cylindrical morphology (a) with one cycle of alumina seed and (b) without an alumina seed layer. Samples were excited at 260nm. Blue-shifted photoemission is present at 335nm (3.35eV) characteristic of 1.5nm – 2nm particle size. Defect state photoemission is visible at 520nm for sample with the alumina seed layer and five cycles of ZnO. This is consistent with the presence of the Raman peak suggesting that this is where full defects states form. For samples with a seed layer and less than five cycles of ZnO we see the growth of defect state emission that is not fully developed. Defect states are not present in samples without a seed layer.

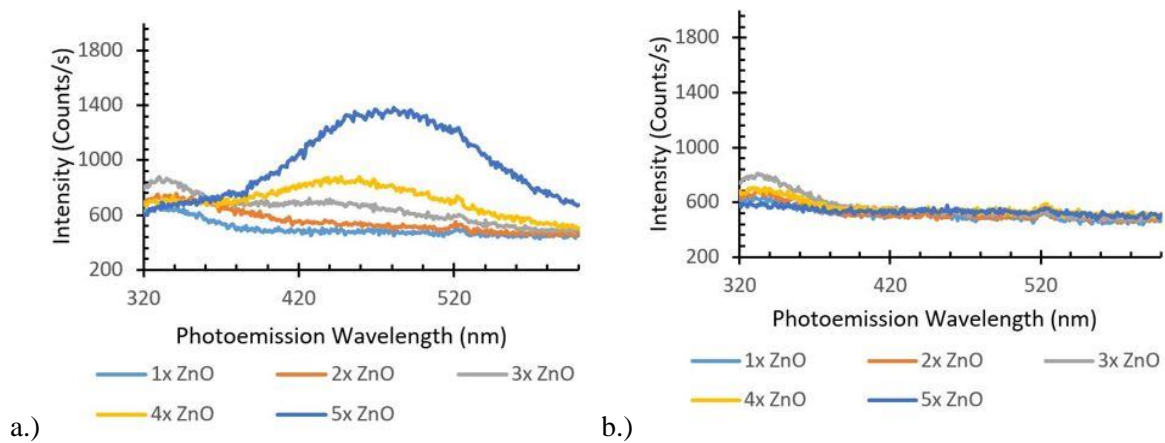

**Figure 7:** Photoluminescence of ZnO infiltrated PS-b-PMMA of spherical morphology (a) with one cycle of alumina seed and (b) without an alumina seed layer. Samples were excited at 260nm. Blue-shifted photoemission is present at 335nm (3.35eV) characteristic of 1.5nm – 2nm particle size. Defect state photoemission is visible at 470nm for sample with the alumina seed layer and five cycles of ZnO. This is different than the defect state emission present in the cylindrical BCP morphology with an alumina seed layer. For samples with a seed layer and less than five cycles of ZnO we see the growth of defect state emission that is not fully developed. Defect states are not present in samples without a seed layer. This is the same results for both morphologies of BCP

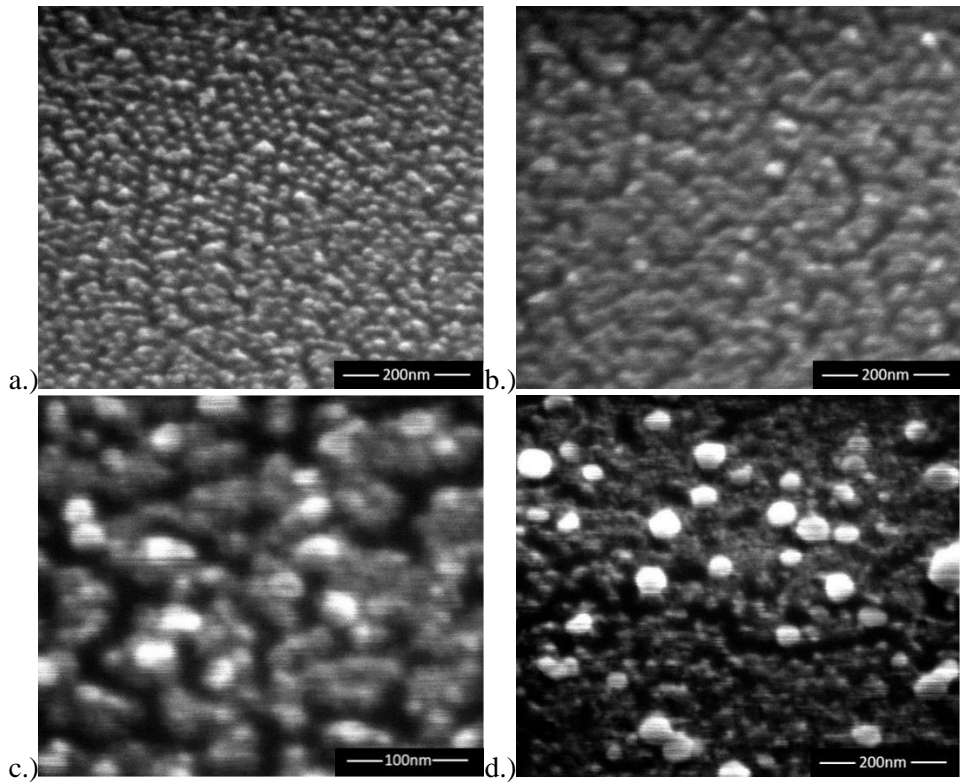

**Figure 8:** SEM images of cylindrical PS-b-PMMA infiltrated with 3x [$H_2O$:DEZ] etched 40s with non-biased oxygen-plasma and annealed at (a) 450 °C, (b) 500 °C, (c) 550 °C, and (d) 600 °C. We can see that samples that are annealed at 500 °C and above begin to coalesce and form larger structures.

Annealing was done in air using a rapid thermal processing (RTP) tool to determine conditions for which our samples could be annealed and still remain undisturbed (**Cheng 2011**). Annealing is an attractive process for ZnO nanostructures because it is well known that annealed ZnO can be used as single photon sources (**Choi 2014; Jungwirth 2015**). Four test samples using cylindrical BCP morphology with 1x [$H_2O$:TMA] and 3x [$H_2O$:DEZ] were prepared and used to test annealing conditions. PL was taken of one of the test samples before annealing. The samples were annealed at 450 °C, 500 °C, 550 °C, and 600 °C, each for 20min. SEM imaging of the samples after each of the anneals was taken to check for any visible changes after annealing. Agglomeration of the ZnO nanostructures was observed starting with the sample annealed at 500 °C. Starting at 550 °C large clusters of ZnO begin forming in bulk. **Figure 8** shows that samples appear to coalesce more as the anneal temperature increased. PL of the samples after anneal showed that when the samples were annealed at 600 °C we observe bulk ZnO emission with a peak around 370nm further supporting the presence of confinement effects for unannealed samples. All PL scans for annealed samples were excited with 260nm light normal to the samples surface with an emission range between 320nm – 600nm. We used a 2nm slit width, 2nm step, and 1s integration time. We performed 31 scans and averaged them all. For our control sample, the averaging began to mask the 335nm emission state due to the severe contrast in intensity of our defect state emission. Annealing appeared to collapse the overwhelming defect state emission, showing standard ZnO emission. The sample annealed at 600 °C showed emission at 370nm consistent with its SEM image that suggested coalescing, as is shown in **Figure 9.**

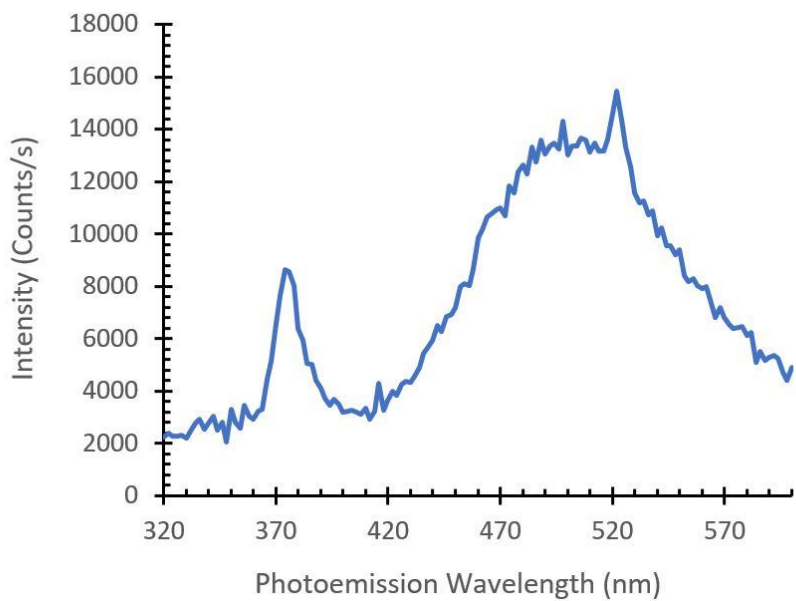

**Figure 9:** Photoluminescence of ZnO infiltrated PS-b-PMMA cylindrical morphology with one cycle of alumina seed layer and three cycles of ZnO after anneal at 600 °C. Emission at 370nm shows bulk ZnO emission, consistent with the SEM seen in Figure 7, showing that annealing at 600 °C causes structures to coalesce into bulk ZnO.

## 3. Summary and Conclusions

We have demonstrated a growth study for nanostructures ZnO on PS-b-PMMA and characterized the properties of these structures based on seed layer and block copolymer morphology. Samples grown with a seed layer have additional defect state photoemission at 470nm and 520nm, for spherical and cylindrical BCP morphology, respectively. Defect states are not present in samples prepared without a seed layer and intensity of photoemission is significantly less. The difference of defect states fluorescence is likely due to the size difference in PMMA domains in the BCP. All samples display photoluminescence at 335nm (3.70eV) which is blue-shifted from bulk ZnO emission at 370nm (3.35eV). This suggests quantum confinement effects which can be related to particle sizes of 1.5nm to 2nm, grown within the 15nm PMMA domains. These ZnO nanostructures are isolated emitters for all samples with less than five cycles of ZnO infiltration. At five cycles of ZnO we begin to see Raman peaks corresponding to the development of a thin film. We have also shown that annealing of samples can cause the nanostructures to coalesce, corresponding to bulk ZnO emission.

Our results show that infiltration synthesized ZnO nanostructures have a significant advantage in size and uniformity. These nanoparticles have characteristic emission and are smaller than other ZnO nanostructures reported in literature. These results suggest that ZnO nanostructures grown via infiltration has the potential be used in a variety of applications, such as in quantum materials and photonics. Observations also imply that pattering of samples pre-infiltration could provide significant advantages in control and placement of these nanostructures, pushing past previous limitations in nanoparticle technology. Examples of preliminary results in this direction are shown in **Figure 10**. By using electron beam lithography with a negative resist, SU-8, and reactive ion etch on a 1x $Al_2O_3$ seeded, 3x [$H_2O$:DEZ] ZnO infiltrated sample, we are able to isolated 2 to 3 single ZnO infiltrated nanospheres on a Si pedestal.

Characterization of these structures are beyond the scope of the current manuscript and will be published at a later time.

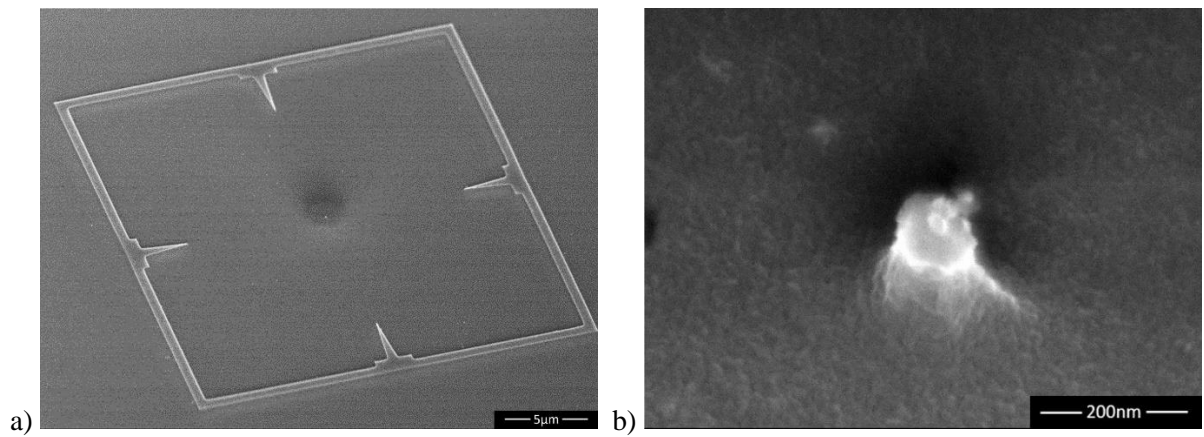

**Figure 10**. Lithographically isolated ZnO infiltrated PS-b-PMMA nanospheres. a) SEM micrograph showing registration marks pointing at an isolated structure in the middle. b) SEM micrograph showing a 100 nm wide Si pedestal with 2 to 3 20 nm infiltrated PS-b-PMMA nanospheres on top.

## 4. Methods

The infiltration process was operated at 95 °C, below the glass transition temperature of PMMA in an Arradiance Gemstar 8 ALD tool. The precursors used were water ($H_2O$), tetra methyl ammonium (TMA), and diethyl zinc oxide (DEZ), all purchased from Stream Chem. Cycles aluminum oxide and zinc oxide were applied with [$H_2O$:TMA] and [$H_2O$:DEZ], respectively. A half cycle of TMA or DEZ consists of a series of pulses of precursor injection for 4 minutes for each of the water, aluminum, and zinc. This is followed by a nitrogen flush for 40s and a 3s chamber pump preceded the next injection. Each half cycle injection includes 80 ms pulses of precursor with 10 s delays in between pulses for each of the water, aluminum, and zinc.

A Nanolog spectrofluorimeter from Horiba Scientific was used for PL measurements. The tool uses a broadband Hg source, with a monochromator, and a separate detection spectrometer. The first measurements taken acquired emission for multiple excitations plotted on a 3D contour plot. This was refined to single excitation of 260nm while recording emission between 320nm to 600nm. Both emission and excitation detectors were dark-current-corrected, and then the data was normalized to the lamp intensity as measured by the excitation detector. The excitation light was normal to the surface of the samples and the emission light was detected at a 22.5° angle with a 305nm high pass filter in front.

Annealing was done using a rapid thermal processor from Annealsys. The anneal process started with a pump down of both pyro chambers for 60 s each. The pressure was then sent to < 0.1 Torr followed by a delay of 25 s. while keeping the main heating chamber > 200 °C. The nitrogen gas was then introduced at 500 R?* over a period of 30s. The temperature was then ramped to 300 °C at 1 °C/s where it then dwelled at for 300 s. The oxygen gas was then introduced at 300 R?* and the nitrogen gas was brought down to 300 R?* over the duration of a minute. The temperature was then ramped to the desired temperature (450 °C, 500 °C, 550 °C, or 600 °C) at 2 °C/s at which point it dwelled for 1200 s before being ramped back down to 2 °C/s.

## 5. Acknowledgments

Use of the Center for Nanoscale Materials, an Office of Science user facility, was supported by the U.S. Department of Energy, Office of Science, Office of Basic Energy Sciences, under Contract No. DE-AC02-

06CH11357. This work was also supported by the University of Chicago Materials Research Center (MRSEC) IRG3-Engineering Quantum Materials and Interactions Contract #2-60700-95, and the University of Chicago Research Experience for Undergraduates (REU) funded by the National Science Foundation (NSF), contract 1420709.